\pdfoutput=1
\documentclass{jinst}
\let\ifpdf\relax 

\usepackage{lineno}
\usepackage{amsmath}
\usepackage{cite} 

\title{\boldmath Performance of active edge pixel sensors}

\author{Marco~Bomben$^1$\thanks{Corresponding author.}, Audrey~Ducourthial$^1$,
Alvise~Bagolini$^2$, Maurizio~Boscardin$^2$, Luciano~Bosisio$^3$, Giovanni~Calderini$^1$, Louis D'Eramo$^1$, Gabriele~Giacomini$^4$, Giovanni~Marchiori$^1$, Nicola~Zorzi$^2$, Andr\'e Rummler$^5$ and Jens Weingarten$^6$.\\

\hspace{-1ex}$^{1}$Laboratoire de Physique Nucleaire et de Hautes \'Energies (LPNHE), Paris, France\\
\llap{$^2$}Fondazione Bruno Kessler, Centro per i Materiali e i Microsistemi (FBK-CMM), 38123 Povo di Trento (TN), Italy\\
\llap{$^3$}Universit\`a di Trieste, Dipartimento di Fisica and INFN, 34127 Trieste, Italy.\\
\llap{$^4$}Brookhaven National Laboratory, Instrumentation Division 535B, Upton, NY - USA; was with Fondazione Bruno Kessler, Centro per i Materiali e i Microsistemi (FBK-CMM), 38123 Povo di Trento (TN), Italy\\
\llap{$^5$}LAPP, CNRS/IN2P3 and Universit\'e Savoie Mont Blanc, Annecy-le-Vieux, France\\
\llap{$^6$}II Physikalisches Institut, Georg-August-Universit\"at, G\"ottingen, Germany
}

\abstract{To cope with the High Luminosity LHC harsh conditions, the ATLAS inner tracker has to be upgraded to meet requirements in terms of radiation hardness, pile up and geometrical acceptance. The active edge technology allows to reduce the insensitive area at the border of the sensor thanks to an ion etched trench which avoids the crystal damage produced by the  standard mechanical dicing process. Thin planar n-on-p pixel sensors with active edge have been designed and produced by LPNHE and FBK foundry. Two detector module prototypes, consisting of pixel sensors connected to FE-I4B readout chips, have been tested with beams at CERN and DESY. In this paper the performance of these modules are reported. In particular the lateral extension of the detection volume, beyond the pixel region, is investigated and the results show high hit efficiency also at the detector edge, even in presence of guard rings.}


\begin{document}
\maketitle
\flushbottom

\section{Introduction}
\label{sec:intro}

Pixel sensors are the standard choice for the innermost layers of charged 
particle tracking and for vertexing  at high energy 
colliders~\cite{AtlasPixels,IBLTDR,CMSPixels}. 
CERN plans to fully exploit the potential of the Large Hadron Collider (LHC) by upgrading it into a high luminosity collider, the High Luminosity LHC~(HL-LHC)~\cite{HL-LHC}. 
To provide the best possible tracking performance together with hermetic coverage, the new pixel sensors for the future HL-LHC ATLAS inner tracker (ITk)~\cite{itk_strips_tdr} must have a very large geometric acceptance, bearing in mind that detector module shingling will be very limited. For example, for the future ITk the distance from the active region  to the cut edge of pixel modules has to be smaller than 100~$\mu m$~\cite{itk_strips_tdr}.

FBK-Trento and LPNHE-Paris produced pixel sensors prototypes characterized by  a reduced dead area at the edge, whose width is compatible with the requirements for the upgrades of the ATLAS tracker~\cite{itk_strips_tdr}.
The joint FBK-LPNHE~\cite{bib:nim2012} planar  production was composed of 200~$\mu m$ thick n-on-p sensors whose boundaries are delimited by an ``active edge''. The active edge is one of the possible choices to realize ``edgeless'' detectors, {\it i.e.} detectors with no (or very limited) insensitive area. Along the sensor border a trench is dug by deep reactive ion etch (DRIE), reaching through the whole thickness of the substrate (hence a support wafer is required).  The  trench is then doped with boron and  filled with polysilicon. The cut realized through DRIE produces an edge region 
much less damaged than the one resulting from a standard diamond-saw cut. This leads to less generation centers hence lower leakage current generated at the border. Moreover, the edge doping prevents the depletion region from reaching the physical trench walls, hence carriers created at the edge  do not experience an electric field, are not effectively separated and just recombine, without contributing significantly to the device leakage current. 
These pixel sensors were intended as a first step 
toward edgeless radiation hard pixel modules; for the latter thinner sensor wafers are needed, to better cope with the high fluences expected at the HL-LHC~\cite{itk_strips_tdr}.

In this paper the performances of pixel detector module prototypes are reported; the modules were composed by a pixel sensor taken from the joint FBK-LPNHE production, bump bonded to a FE-I4B readout chip~\cite{FEI4}.
The detectors were evaluated on beam; the main characteristics of the tested detectors are reported in Section~\ref{sec:device_charact}. In Section~\ref{sec:exp_setup}, the experimental setup will be presented, including the beam line and the tracking telescope, the algorithms for track reconstruction and data analysis. 
 Beam test results, including hit and charge collection efficiency, and spatial resolution will be presented in Sections~\ref{sec:unirr_results}; in particular the 
efficiency at the detectors edge will be discussed.
Finally (Section~\ref{sec:conclusions}) conclusions will be drawn and  future plans will be presented.

\section{Devices under test}
\label{sec:device_charact}
\subsection{Description of tested devices}

\begin{figure}[!htb]
\centering
\includegraphics[width=0.30\textwidth]{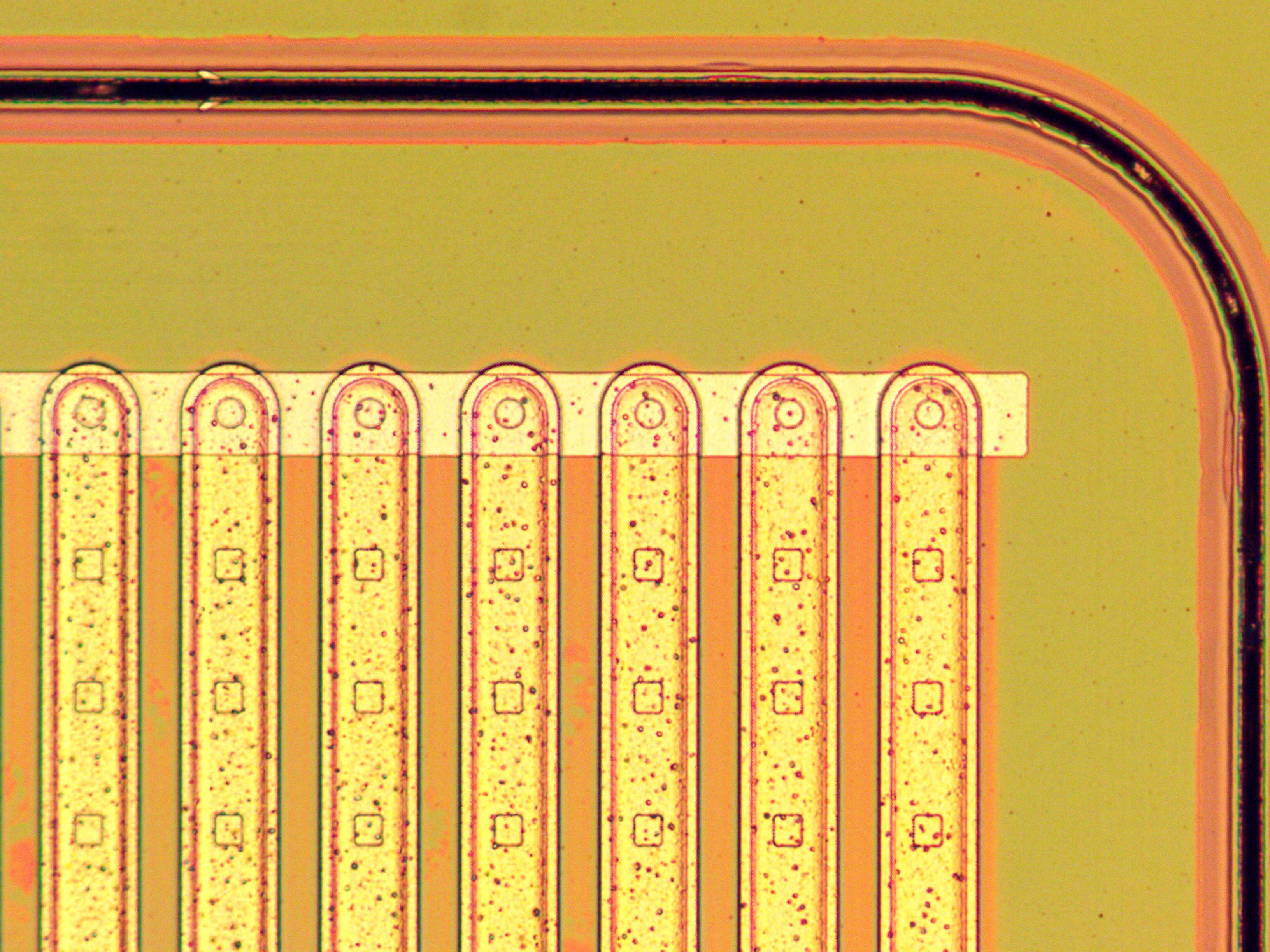}
\includegraphics[width=0.30\textwidth]{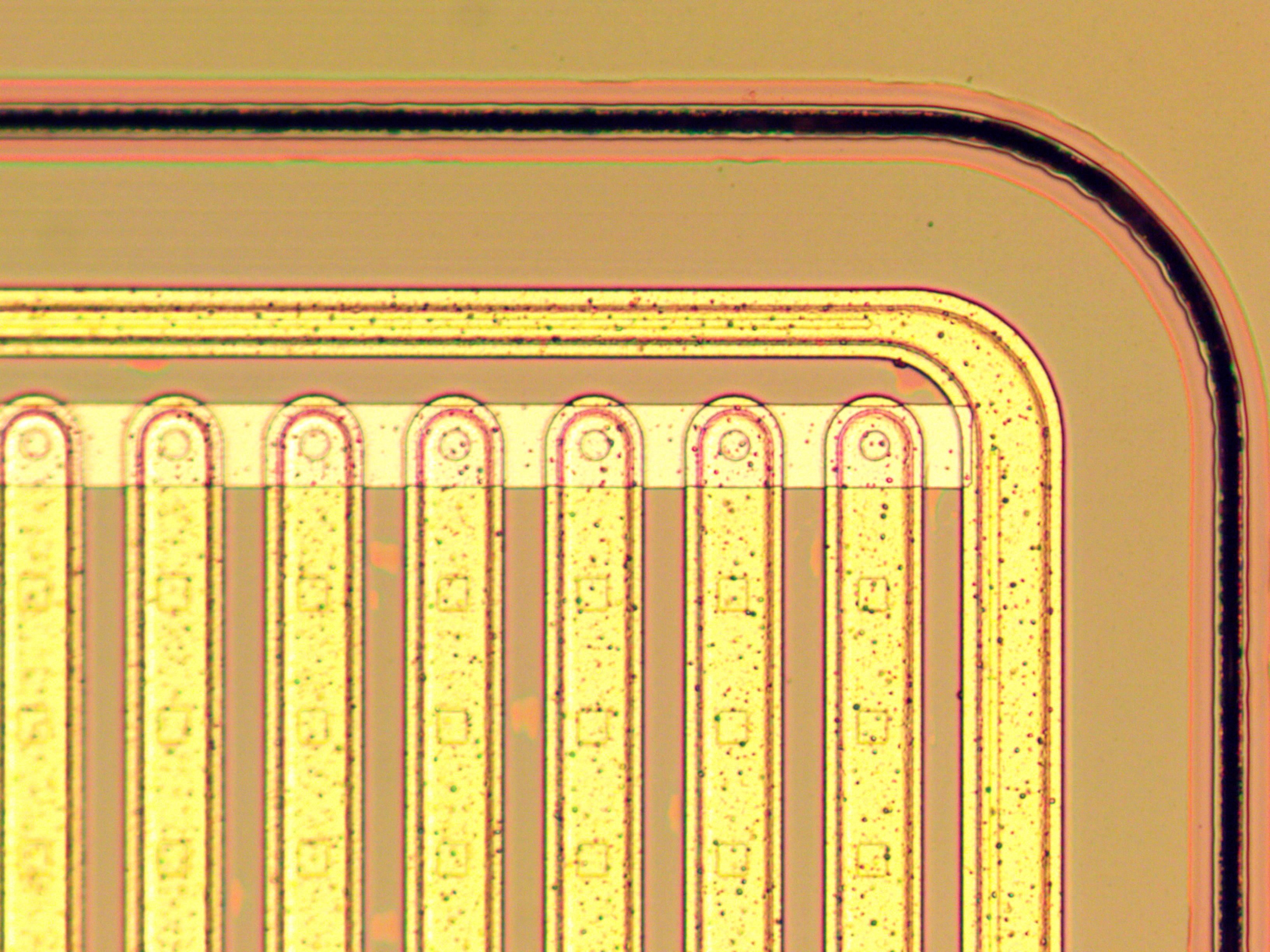}
\includegraphics[width=0.30\textwidth]{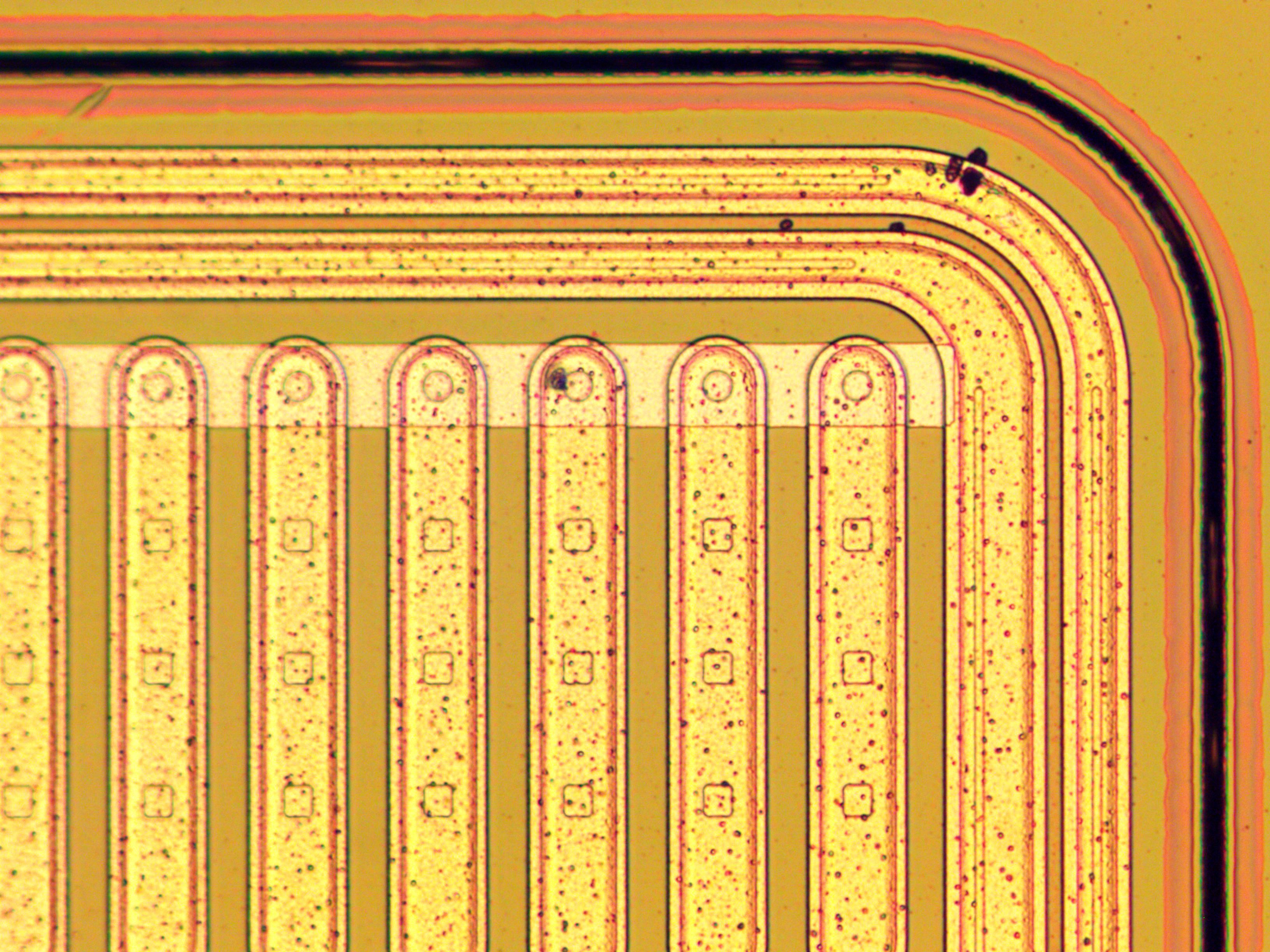}
\caption{\label{fig:lpnhe5_4_7_pic}Microscope picture of corners of the  (left) LPNHE5, (middle) LPNHE4  and (right) LPNHE7 sensor. The black line at the top and on the right is the trench. The  shortest distance from the pixels to the trench is 100~$\mu m$ for all the three sensors. For LPNHE4 there is  one GR surrounding the pixel matrix; for LPNHE7 there are two GRs. The pictures show also a temporary metal strip~\cite{bib:metal} shorting the pixels: it was used at wafer level for checking the sensor current  but it was removed from the detectors tested in this work.}
\end{figure}

Three sensors were bump-bonded to FE-I4B readout chips at IZM Berlin\footnote{Fraunhofer-Institut f\"ur Zuverl\"assigkeit und Mikrointegration IZM - Gustav-Meyer-Allee 25, 13355 Berlin, Germany}. Each pixel sensor is composed of  336~rows~$\times$~80~columns of rectangular pixels cells whose dimensions are 50~$\mu m$~$\times$ 250~$\mu m$.
The main difference among the three sensors is  the number of guard rings (GRs) surrounding the active area, ranging from zero to two. In Figure~\ref{fig:lpnhe5_4_7_pic} a detail of the 
sensor edge can be seen for all the three samples.  

LPNHE5 has no GRs, LPNHE4 has one GR and LPNHE7 has two GRs. All sensors are 200~$\mu$m thick n-on-p and include a uniform p-spray implant on the pixels side to provide enough insulation among them. LPNHE4 and LPNHE5 sensors have, in addition, p-stops implants that surround the implants of pixels and GRs. 
The main  characteristics of the devices are summarized in Table~\ref{tab:device_charact}.

\begin{table}[htbp]
\centering
\caption{\label{tab:device_charact}Tested devices characteristics.}
\smallskip
\begin{tabular}{|l|c|c|}
\hline
Name&Number of GRs& p-stop implant\\
\hline
LPNHE5 & 0 & yes\\
LPNHE4 & 1 & yes\\
LPNHE7 & 2 & no\\
\hline
\end{tabular}
\end{table}

This paper covers results from the LPNHE5 and LPNHE7 devices. 
 The LPNHE4 module was used in an irradiation experiment before the beam tests. Laboratory measurements after irradiation showed that, due to the lack of  electrical insulation layer 
between the sensor and the FEI4-B readout chip (for a discussion of this issue see for example~\cite{rossi2006pixel}), it couldn't be biased up to full depletion. Hence there won't be results for irradiated detectors from this pixel sensors production.
 

During all measurements the innermost GR, if present, was kept at 
 ground voltage by the FE-I4B readout chip; the second GR, when present, was left floating. The depletion voltage for all three devices was about 20~V.

\begin{figure}[!htb]
\centering
\includegraphics[width=0.65\textwidth]{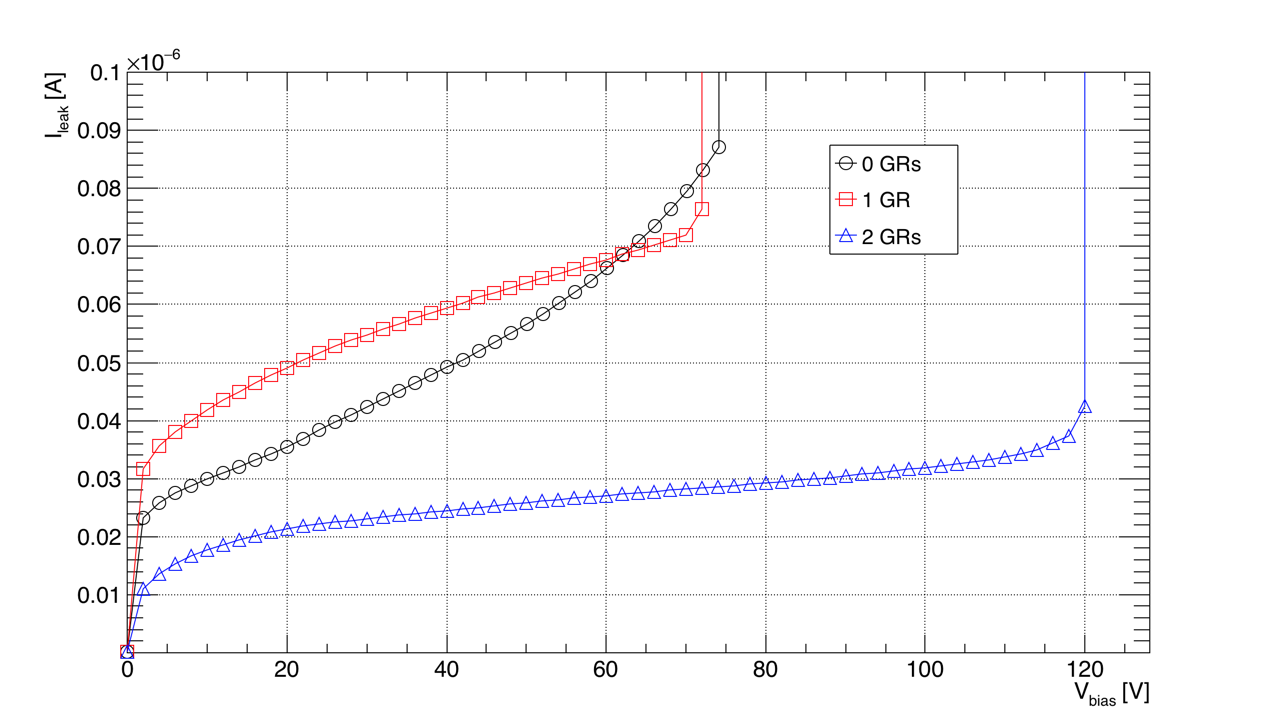}
\caption{\label{fig:IV_GRs}Current-Voltage curves for 
 test structures featuring different number of GRs. The 
innermost GR, if present, was kept at ground voltage. The  shortest distance from the pixels to the trench is 100~$\mu m$. The measurement for the test structure with 2 GRs was taken at a lower temperature with respect to the other two samples.}
\end{figure}

The effect of GRs on the breakdown voltage can be seen in Figure~\ref{fig:IV_GRs}, 
where the current-voltage curves of test structures featuring FEI4-like pixels and different number of GRs are reported;  the distance between the last pixels and the doped trench is 100~$\mu m$. These test structures come from the same wafer of the sensors tested on beam.   The breakdown voltage increases by more than 70\% (from 70 to 120~V) by adding a second, floating GR.

\subsection{Detector configuration}
\begin{figure}[!htb]
\centering
\includegraphics[width=0.65\textwidth]{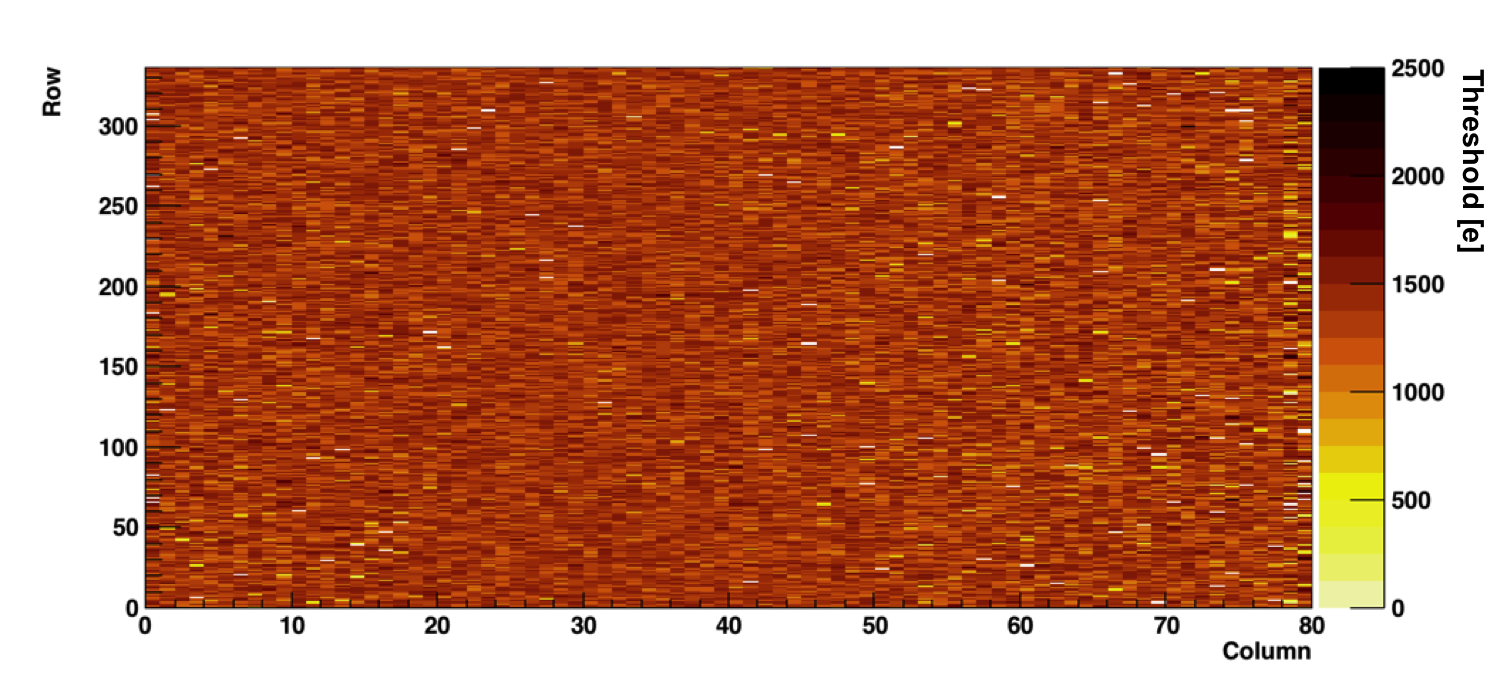}
\caption{\label{fig:threshold}Pixel threshold values for LPNHE7. On the abscissa is the pixel column index, on the ordinate axis is the pixel row index.
The tuning target value was 1400~e; the sensor bias voltage was 40~V.}
\end{figure}

Before laboratory and beam tests the threshold and gain settings of the readout electronics are carefully tuned. When choosing the threshold, a compromise has to be found between a high threshold, which decreases the number of noise hits but decreases the signal efficiency as well, and a low threshold, with opposite effects. For our detectors, a typical threshold is 1400~e, which corresponds to a tenth of the expected most probable value signal amplitude due to a minimum ionizing particle (MIP) crossing the sensor at normal incident angle. A typical result from  threshold tuning\cite{FEI4,USBpix} can be see in Figure~\ref{fig:threshold}; the threshold dispersion is of the order 200~e. The signal amplitude in the sensor is measured in  units of Time over Threshold (ToT): a clock counts when the shaped signal  goes above threshold and stops when the signal falls below threshold; the difference between those two crossings is the ToT~\cite{FEI4}. During the tuning of the electronics, the correspondence between ToT value and input charge is calibrated.\\

\section{Experimental setup}
\label{sec:exp_setup}
\subsection{Beam lines, telescopes and data acquisition systems}
The results presented in this publication are based on data 
taken at the DESY beam test facility\footnote{http://testbeam.desy.de/} and at the CERN 
North Area experimental area\footnote{http://sba.web.cern.ch/sba/}. 
At DESY 4~GeV/c momentum electrons were used; the beam was almost continuous. At CERN 120~GeV/c momentum positive pions were used; the time structure of the 
beam was organized in spills within a super cycle of a few tens of 
seconds. 

At both laboratories the data were recorded using a copy of the Eudet/AIDA telescope~\cite{Jansen2016}. This generation of beam telescopes consists of six detection planes equipped with the  Mimosa26~\cite{mimosa26} monolithic active pixel sensors, with a pitch of 18.4~$\mu$m. 
The data read out was triggered by the coincidence of plastic scintillators,
whose area was of about 1~cm$^2$. 
The data from the DUTs were recorded using two different Data Acquisition (DAQ) systems: the Reconfigurable Cluster Element (RCE)~\cite{RCE} system and the UsbPix~\cite{USBpix} system.
The typical averaged\footnote{Averaged over a supercycle at CERN} trigger rate was in the range of 250-1000~Hz, depending on the 
beam conditions and on the DAQ system used for the devices under test (DUTs).

The DUTs were located between the two arms of the telescope (each arm having three detection planes). To screen the DUTs from the light, they were operated inside a cooling box, capable  of maintaining the DUT temperature constant.

\subsection{Analysis process and measured observables}
The track reconstruction consists of a set of algorithms, implemented in the EUTelescope framework~\cite{eutelescope}, to process raw data into tracks.

After the data taking, as a first step a noisy pixels data bank is created both for telescope planes and DUTs, looking at pixels which fired at a frequency higher than a certain threshold; at later stages, signal from the pixels appearing in the data bank are discarded. Next comes  the clustering step: in each plane neighboring pixels firing in the same bunch crossing are grouped together to form clusters. For each cluster, hit coordinates are computed in the global frame and a first alignment of the telescope planes and the DUTs is performed. The final alignment, based on the Millipede algorithm~\cite{Millipede}, is then performed to align each DUT plane independently from other DUT planes.
Eventually, tracks are reconstructed using a  Kalman-filter based algorithm and a $\chi ^2$ fit is performed to obtain the best possible track parameters with hits on each plane.
At the end of the process a ROOT~\cite{root} file is created containing basic observables  ready to be analyzed in the data analysis framework, TBmon2 software~\cite{tbmon2}. 
TBmon2 allows studying the quantities discussed below. 

\subsubsection*{Global, in-pixel and edge hit efficiency}

\begin{figure}[htbp]
\centering 
\includegraphics[width=.45\textwidth,origin=c,angle=0]{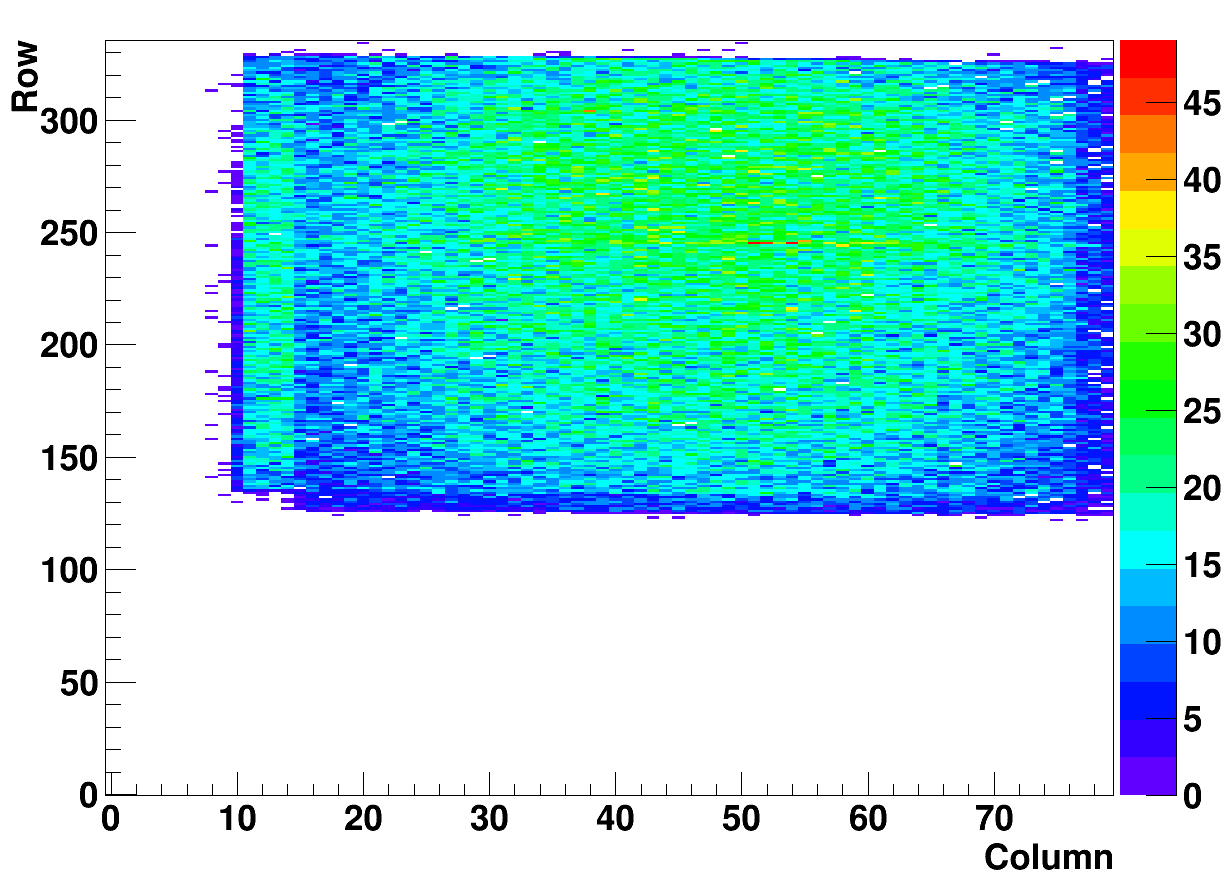}
\qquad
\includegraphics[width=.45\textwidth,origin=c,angle=0]{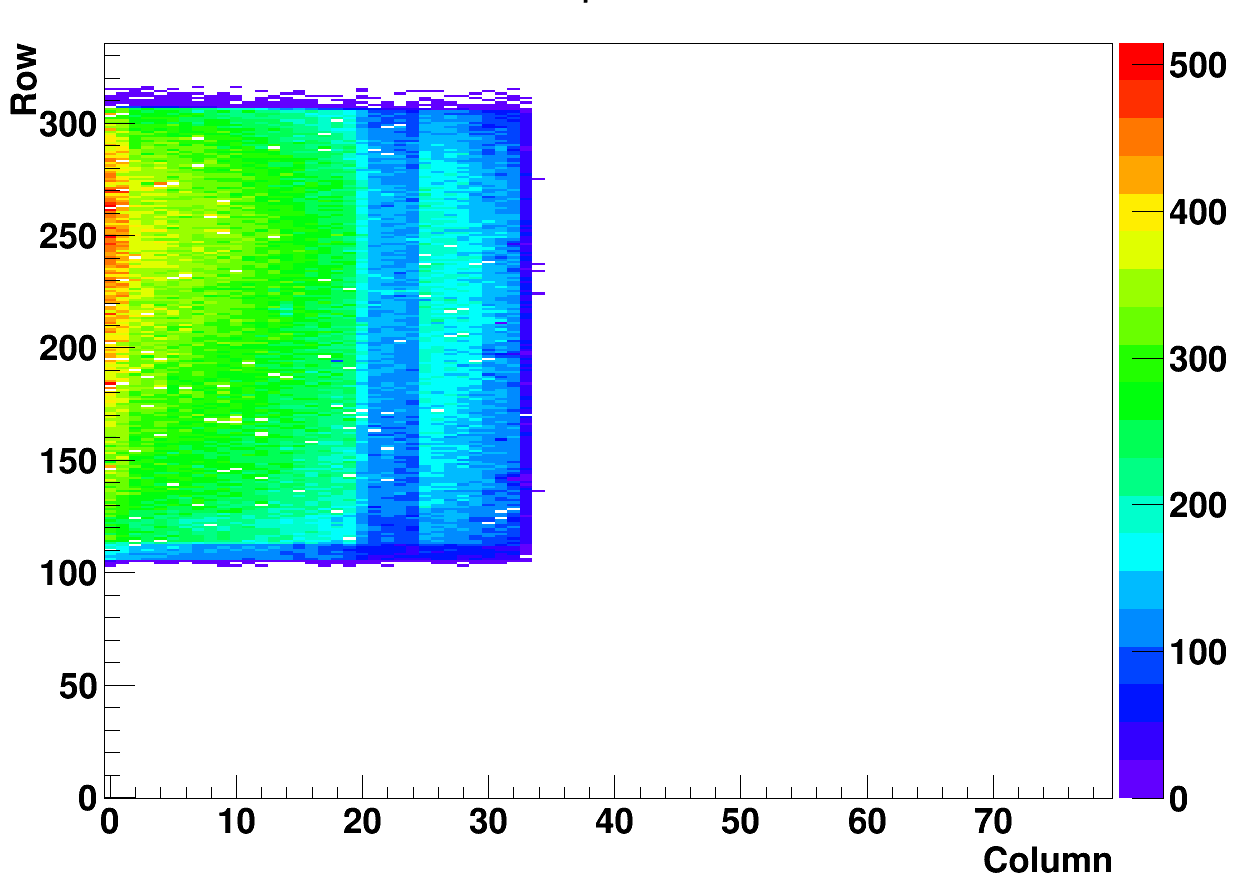}
\caption{\label{fig:hitmap} Hit map of a tested sensor in beam. On the abscissa is the pixel column index, on the ordinate axis is the pixel row index. (Left) the beam is focused on the center of the sensor; (right) the beam  is focused on the edge, which allows to perform edge efficiency scan. The area where hits are seen is a 1 $ cm^2$ rectangle and correspond to the area of the trigger scintillator.}

\end{figure}

The global hit efficiency is defined as the  fraction of reconstructed tracks crossing a sensor that have an associated hit in that sensor. A bad bump bonding can degrade severely the efficiency of the sensor. 
The quoted efficiency is measured in a fiducial region, defined by the surface of the pixel module where each pixel cell is 
hit by at least 1 track. From Figure~\ref{fig:hitmap} it can be seen that the fiducial region,  defined by the trigger scintillators area, is smaller than the surface of the detector. Nonetheless 
the uniformity in threshold show in Figure~\ref{fig:threshold} is a good 
indication that the performance measured in the fiducial area can be taken as valid also outside it, 
hence the hit efficiency be interpreted as global.

The in-pixel hit efficiency is obtained by superimposing the 2D maps of efficiency as a function of the local position in each pixel cell of the sensor, the granularity of this analysis being of the order of the total pointing resolution (sum of the telescope resolution and the multiple scattering average shift). The in-pixel efficiency gives valuable information on the homogeneity of the charge collection, stressing the presence of low efficiency areas due, for instance, to permanent biasing structures. Our sensors do not include permanent biasing structures, since for testing purposes they are polarized thanks to a temporary metal line~\cite{bib:metal}, which is then removed before bump bonding.

To assess whether the active edge ensures a high hit efficiency in the  area between the last pixels and the doped trench, an efficiency measurement as a function of the track position in the edge area is performed, using data collected with the beam focused on the edge area; see also Figure~\ref{fig:hitmap}.

The impact of the GRs on the efficiency is studied by comparing numerical device simulations with the edge hit efficiency profiles. The lateral depletion can be investigated looking at the edge efficiency performance
 for several values of the bias voltage.

\subsubsection*{Hit residuals and spatial resolution}
The hit residuals are defined as the difference between hit position in the sensor and the position of the intersection between the associated reconstructed track and the DUT. 
The study of the residual distribution gives valuable information on the sensor spatial resolution after accounting for the pointing resolution of the telescope, multiple scattering and charge sharing between neighboring pixels.\\ The multiple scattering at CERN SPS has a  significantly smaller effect compared to the detector resolution as beam particles are high momentum pions of 120 GeV/c. The spatial resolution is obtained from the RMS of the residual distribution for all clusters. The main components of the clusters residuals distribution are: \begin{itemize}
\item The residual distribution of one-pixel clusters.
This distribution is expected to be flat and to span over a width compatible with the pixel implant one (36 $\mu m$ in the short pixel side). However, since the pointing resolution of the telescope smear the edges of the flat distribution, the residuals can be fitted by a flat distribution convoluted with a Gaussian, whose width gives an estimation of the telescope pointing resolution, convoluted with the multiple scattering induced shift~\cite{stefano}.
\item The residual distribution for two-pixels clusters. The charge sharing occurs in an area between two pixels which is narrower than the pixel pitch, consequently the spatial resolution for a two-pixels cluster is better than for a one-pixel cluster. The distribution is fitted with 2 Gaussians: a narrow one which is the true residual distribution for two-pixels clusters and a broad outlier Gaussian  which takes into account badly reconstructed hits. The RMS of the narrower Gaussian gives an estimation of the spatial resolution for two-pixels clusters. The area of the narrow Gaussian over the area of the sum of the two Gaussians is the fraction of correctly reconstructed two-pixels clusters.
\end{itemize}

\section{Results for tested detectors}
\label{sec:unirr_results}
\subsection{Global Hit Efficiency}
\begin{figure}[htbp]
\centering 
\includegraphics[width=.65\textwidth,origin=c,angle=0]{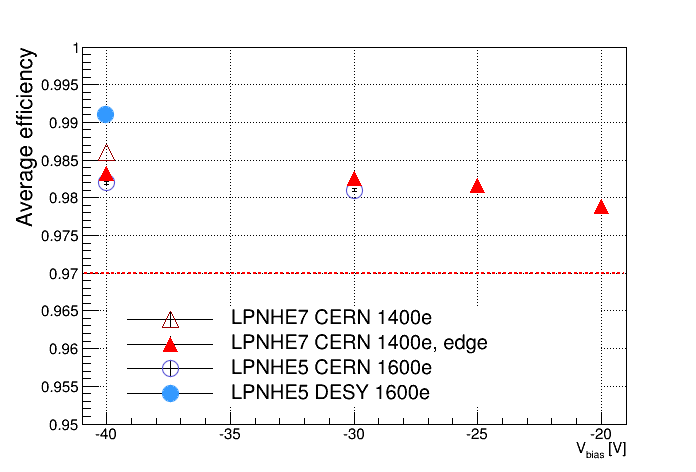}

\caption{\label{fig:eff} Global hit efficiency for the 2 sensors (LPNHE7 and LPNHE5), for various bias points, threshold configurations (1600 e or 1400e ) and beam tests (CERN or DESY). ``Edge'' identifies data taken when the 
beam was focused at the detector periphery.}

\end{figure}

The hit efficiency has been investigated at CERN SPS and DESY with a set of two thresholds corresponding to an input charge of 1400 electrons or 1600 electrons and for various bias points. 
The global hit efficiency is higher than 97.5 $\%$ for both the LPNHE5 and LPNHE7 sensors, as shown in Figure~\ref{fig:eff}.
For LPNHE7 at the CERN SPS with a threshold of 1400 electrons, two beam configurations were investigated, one with the beam focused on the center of the sensor (open triangles), the other with the beam focused on the edge of the sensor (full triangles).
Biasing the sensor above 25~V allows the sensors to reach a 98~$\%$ efficiency whatever the threshold.
\subsection{In-Pixel Hit Efficiency}

\begin{figure}[htbp]
\centering 
\includegraphics[width=.7\textwidth,origin=c,angle=0]{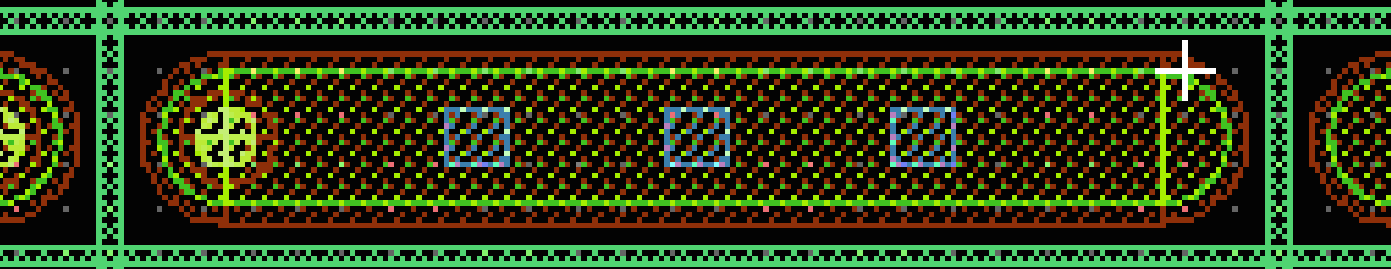}

\includegraphics[width=.75\textwidth,origin=c,angle=0]{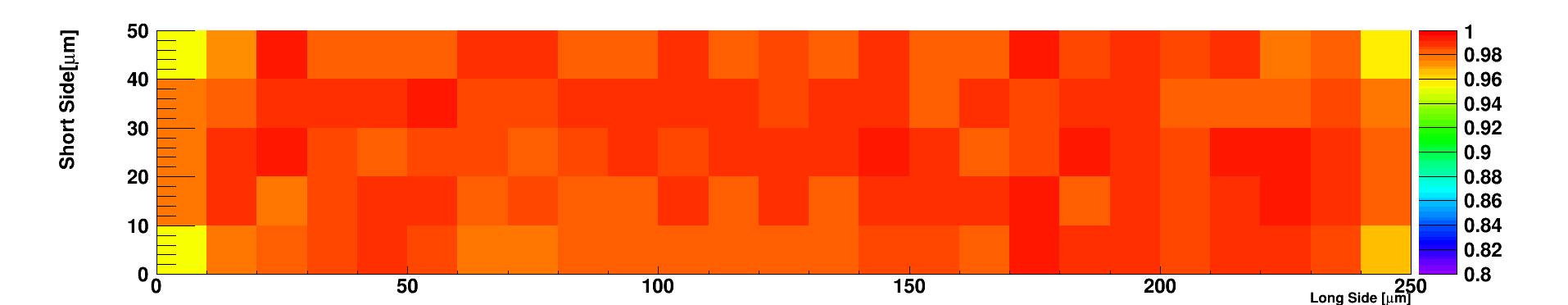}
\caption{\label{fig:pix} Pixel scheme (top) with inner structures: n$^+$-implant, metal contacts, bump bond pad, p-stop... and in pixel efficiency (bottom) for LPNHE7 at 40~V.}

\end{figure}

As observed in Figure \ref{fig:pix}, the in-pixel efficiency is very homogeneous. This high homogeneity shows the interest of using a temporary metal to bias the sensors for electrical tests before bump-bonding instead of adding a permanent structure such as punch-through bias dots. A tiny drop of efficiency can be observed at the pixel corner, where it decreases to  95$\%$. This is due to the charge sharing occurring between 3 or 4 neighboring pixels. In those clusters, the charge induced in one of the pixels could be under threshold and then not taken into account, which biases the hit reconstruction and the hit efficiency.

\subsection{Edge Efficiency}

\begin{figure}[!htbp]
\centering 
\includegraphics[width=0.75\textwidth,origin=c,angle=0]{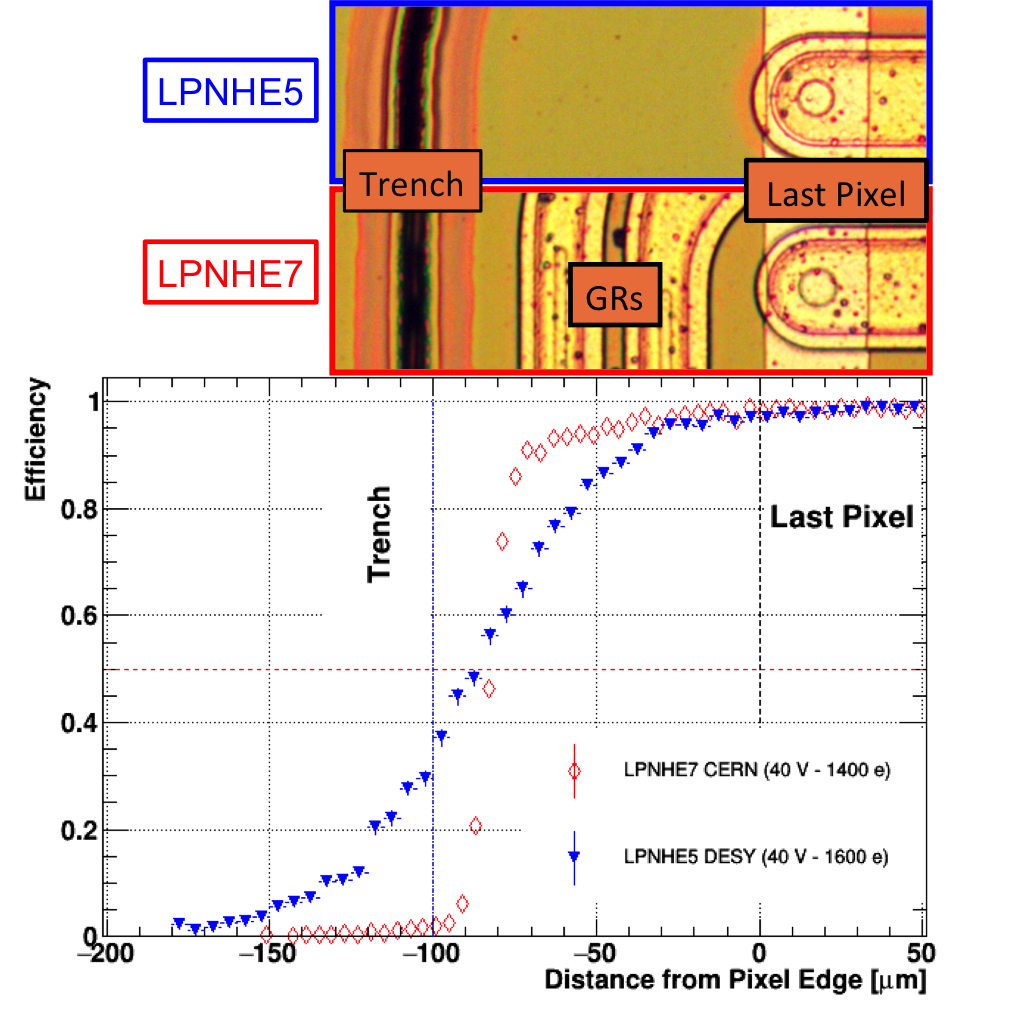}
\caption{\label{fig:edge_comparison}
Edge efficiency profiles for  LPNHE5 (no GRs - full markers) and LPNHE7 (2 GRs - open markers). Laboratory were the data were taken, device bias voltage and threshold are indicated too. The horizontal dashed line marks the 
50\%-point efficiency. The devices photograph on top helps in visualizing which physical area of the pixel is related to the efficiency profile.}
\end{figure}

The hit efficiency at the detector edge for both LPNHE5 and LPNHE7 is presented in Figure \ref{fig:edge_comparison}. LPNHE5 and LPNHE7 were measured at DESY and at CERN respectively; the threshold was set to 1600~(1400)~e for LPNHE5 (LPNHE7), while the bias voltage was 40~V for both detectors. 

Thanks to the active edge technology both detectors are efficient even in the un-instrumented area: for both LPNHE5 and LPNHE7 the efficiency is higher than 50$\%$ up to about 90 $\mu m$ away from the last pixel, that is only 10 $\mu m$ from the cut edge. This performance meets the specifications of ATLAS ITk pixel modules~\cite{itk_strips_tdr} in terms of distance from the active region to the cut edge.

As a reminder, LPNHE7 has 2 GRs, one connected to ground laying between 13~$\mu m$ and 50~$\mu m$ from the last pixel, one floating between 55~$\mu m$ and 80~$\mu m$; LPNHE5 has no GRs. 
The behavior of the 2 samples is rather similar in the first 30~$\mu m$, where the efficiency is basically flat. Then the efficiency  drops faster  for LPNHE5, while  for LPNHE7 the efficiency is a plateau between 0 and -50~$\mu m$ then it smoothly decreases to reach 90~$\%$ at -80~$\mu m$, before sharply dropping to 0. 

Even if data taking conditions were different and clearly sub-optimal for LPNHE5 (higher threshold, multiple scattering,~...), the detector is still quite efficient in the edge area. In particular, it is to be noted that the slope of the hit efficiency curve is consistent with the smearing in the telescope tracking resolution due to the multiple scattering. Nevertheless, further tests on active edge sensors without GRs are necessary, with better experimental conditions. 

For LPNHE7, the good performance in terms of efficiency in the edge area indicates that the presence of GRs does not degrade too much the  hit efficiency, even in the area of the innermost connected GR.

\begin{figure}[!htbp]
\centering
\includegraphics[width=0.497\textwidth,origin=c,angle=0]{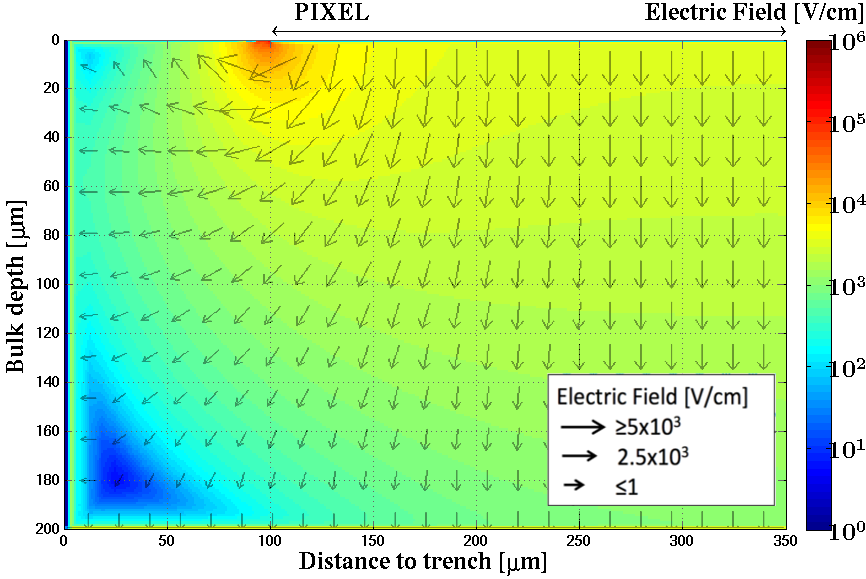}
\includegraphics[width=0.497\textwidth,origin=c,angle=0]{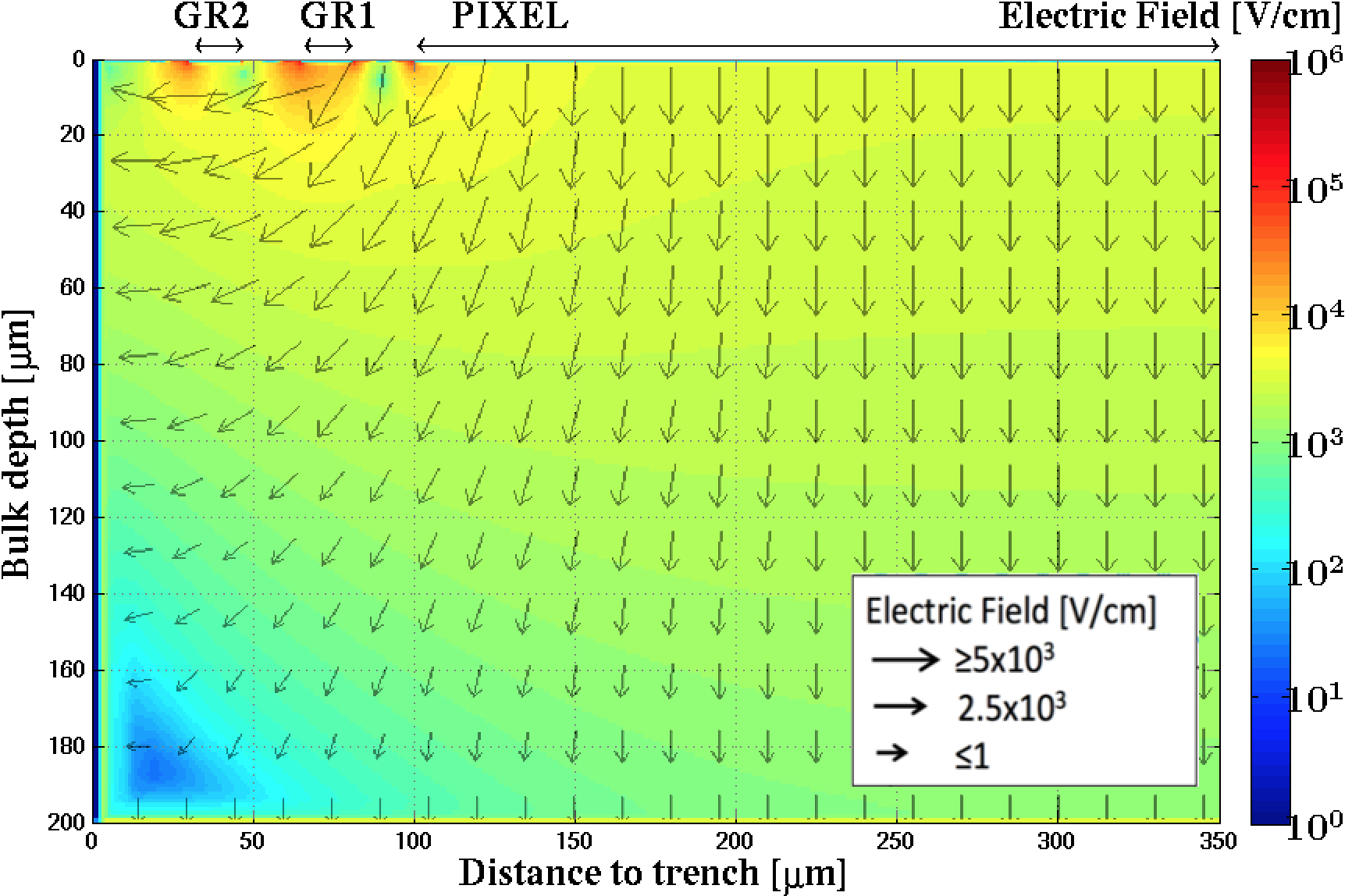}
\caption{\label{fig:EFtcad_edge}Numerical simulation  of the electric field. Left: 0 GRs; right: 2 GRs. The simulated bias voltage value was 40~V.}
\end{figure}

To better understand the efficiency in the GRs region, two dimensional numerical simulations (for details see~\cite{bib:nim2012}) were run; the edge area of  sensors with 0 and 2 GRs and 
a 100~$\mu m$ distance between the last pixel 
and the doped trench were studied. The results are shown in Figure~\ref{fig:EFtcad_edge} for a simulated bias voltage value was 40~V. 

 From Figure~\ref{fig:EFtcad_edge} it can be seen that the GRs do not  deeply influence the electric field lines. The charge carriers, following the electric field lines, are collected by the last pixels if they are electrons or by the trench or backside if they are holes. This seems to be the case from the simulation results, except for electrons generated within a small depth below the GRs. This picture is consistent with the efficiency
 results shown in Figure~\ref{fig:edge_comparison}. 

From Figure~\ref{fig:EFtcad_edge} it can also be  seen that the depleted area is slightly larger for the sensors with 2 GRs and extends till the sensor edge: the GRs are contributing to the depletion of the 
 sensor bulk. 
The simulated electric field magnitude in Figure~\ref{fig:EFtcad_edge} shows a weak electric field region in the bottom left corner; this is due to the presence of two close equipotential planes, the doped trench and the sensor backside. Carriers generated here drift so slowly that they do not produce a signal during the useful integration time of the read-out electronics, and the efficiency drops.

 In summary, based on the above results, supported by numerical  simulations, it can be stated that GRs do not preclude the possibility to have edgeless detectors; their presences make possible at the same time high hit efficiency at the detector edge, by extending laterally the depletion region, and high breakdown voltage (as shown in Figure~\ref{fig:IV_GRs}).

\begin{figure}[htbp]
\centering 
\includegraphics[width=.65\textwidth,origin=c,angle=0]{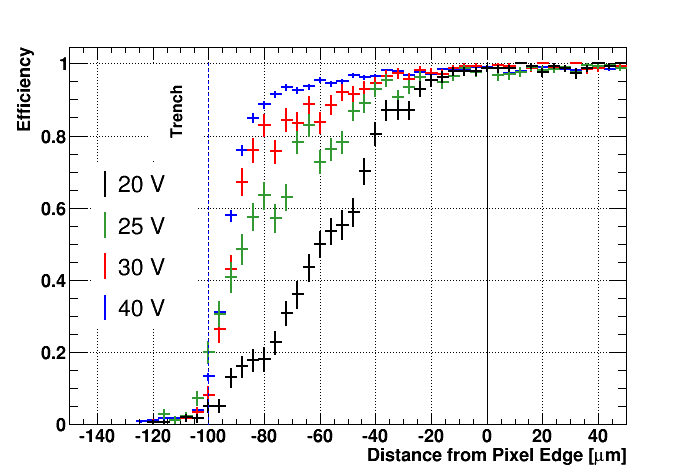}
\caption{\label{fig:comparison2} Comparison of edge efficiency profile of LPNHE7 for several bias voltages}
\end{figure}

In order to further investigate the lateral depletion of the LPNHE7 sensor in the un-instrumented area between the last pixel and the trench, the hit efficiency   was measured as a function of the track distance from  the edge  for several values of the bias voltage, as shown in Figure~\ref{fig:comparison2}.

The edge efficiency is highest at 40~V, where the lateral depletion is such that the efficiency exceeds 50$\%$ up to a distance of 90~$\mu m$ from the pixel edge. At 20~V, the lateral depletion is clearly not completed as the 50$\%$ efficiency point is reached at 60 $\mu m$. The 30~V efficiency profile is quite close to the 40~V curve, although the high efficiency (>95\%) in the region between 50~$\mu m$ and 70~$\mu m$ is possible only at the 40~V. A few events yield non zero efficiency up to 20~$\mu m$ beyond the edge. This is consistent with the spatial resolution of the hits formed by one pixel cell.


\subsection{Spatial resolution}
\begin{figure}[!htbp]
\begin{minipage}[c]{0.1\textwidth}

  \end{minipage}\hfill
  \begin{minipage}[c]{0.5\textwidth}
    \includegraphics[width=0.9\textwidth]{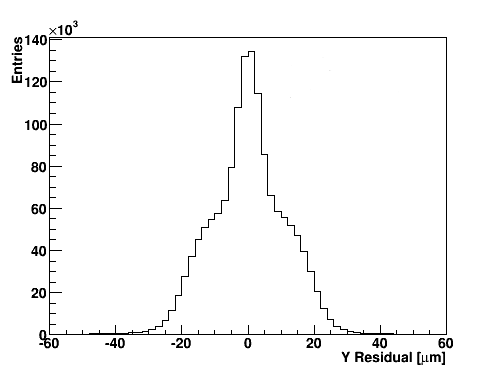}
  \end{minipage}\hfill
  \begin{minipage}[c]{0.4\textwidth}
    \caption{ \label{fig:res_all}Residual distribution of LPNHE7 for all clusters in the short pixel direction (50 $\mu$m pitch). The RMS of the residual is about 11.5 $\mu$m } 
  \end{minipage}\hfill
 
\end{figure}

By looking at the RMS of the cluster residuals reported in Figure \ref{fig:res_all} (data taken at CERN), the spatial resolution in the short direction of the  pixel can be evaluated to be $\sim$~11.5~$\mu m$. This is better than the expected digital resolution for 50~$\mu m$ pitch sensors, i.e. 50$\mu m$ / $\sqrt[]{12}$ $\simeq$ 14.4 $\mu m$. The main reason is of course the presence of clusters formed by two pixels.

\begin{figure}[!htbp]
\centering 

\includegraphics[width=.45\textwidth,origin=c,angle=0]{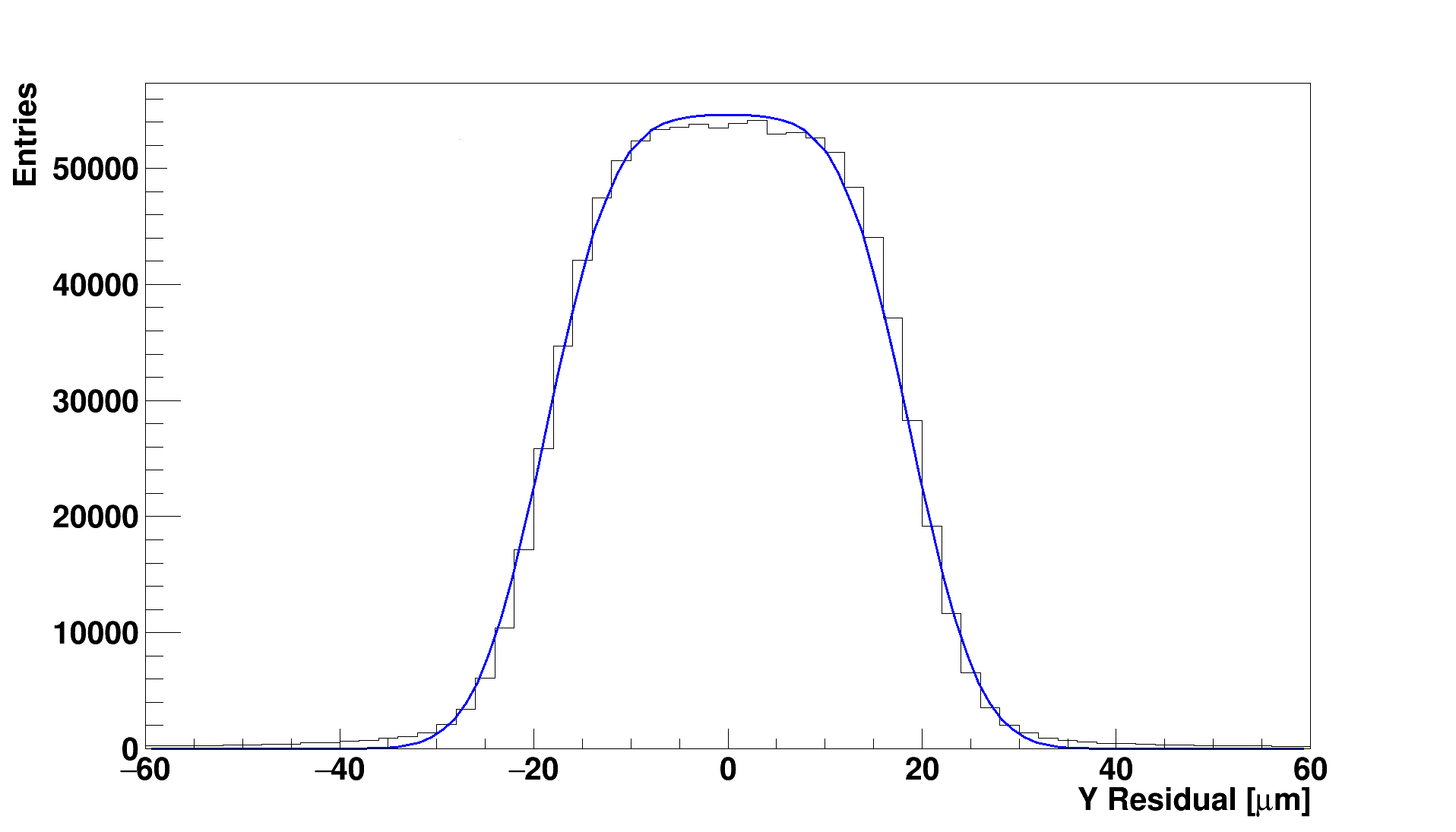}
\includegraphics[width=.45\textwidth,origin=c,angle=0]{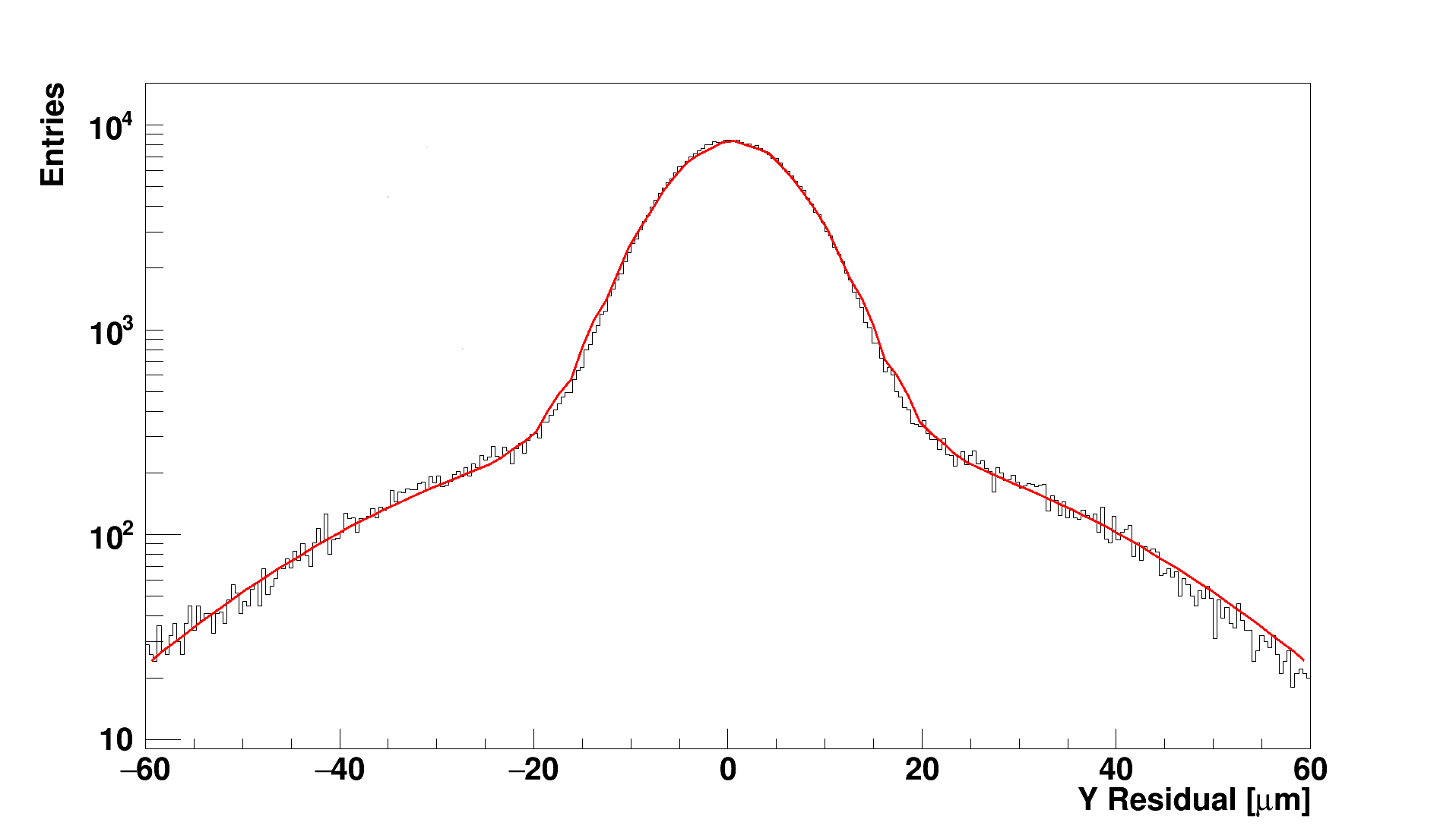}

\caption{\label{fig:res}   Left: residual distribution for clusters of 1 pixel cell fitted with a box function convoluted with a Gaussian. Right: residual distribution in logarithmic scale  of two pixels clusters,fitted with the sum of two Gaussians.}

\end{figure}

The two histograms in Figure~\ref{fig:res} show respectively the residual distribution in the narrow pixel direction for one and two pixels clusters. The RMS of the residuals for clusters of one pixel is of the order of 14~$\mu m$ which is compatible with the expected digital resolution. The distribution is fitted with a box function convoluted with a Gaussian; the pointing resolution of the telescope (convoluted with the multiple scattering effect due to the cooling box for the DUTs) which is obtained by looking at the RMS of the Gaussian, is of the order of 5.5~$\mu$m. \\The residuals distribution of clusters of two pixels is fitted by a convolution of a narrow core Gaussian and a broad outlier Gaussian, the latter to account for badly reconstructed hits. From the fit, the percentage of correctly reconstructed hits is 86~$\%$ and the width of the charge sharing region, given by the RMS of the core Gaussian, is of the order of 7.8~$\mu$m. From those plots it is clear that optimizing the number of two-pixels clusters significantly improves the spatial resolution.

\section{Conclusions and Outlook}
\label{sec:conclusions}
The HL-LHC conditions demand for a completely new tracker for the ATLAS 
experiment. 
To fully exploit the dataset expected at the end of the HL-LHC, the new detector has to be placed as close as possible to the interaction point, which poses severe constraints on the new tracker structure. In particular, the possibility of  shingling  the pixel modules  is very limited, especially along the beams direction, which imposes limits on the insensitive area at the detector periphery. 

In this work it was shown that the active edge technology allows a drastic reduction of the  dead area at the detector periphery. The doped trench at the detector edge allows the depleted area to extend almost to the border of the silicon sensor, without drawing any current from the edge, 
and making it possible to have a hit efficiency higher than 90$\%$ up to 80 $\mu$m from the last pixel cell, 
hence
assuring very high hit efficiency almost everywhere in the detector volume. 
It was also shown that the presence of guard rings does not degrade the hit  efficiency; on the contrary, guard rings help the lateral extension of the depleted region and don't interfere severely with charge collection, making it possible at the same time to achieve a high hit efficiency in the sensor edge area and fairly large operation voltages.

New planar pixel productions exploiting the active edge technology are under development at FBK-Trento, in collaboration with LPNHE-Paris and INFN-Italy. The goal is to reduce the sensor thickness, to better cope with the radiation damage, to further reduce the size of the insensitive edge area and to have smaller pixels for better performance at higher particle rates. 

\acknowledgments
The fabrication of the detectors was supported by INFN. 
This project has received funding from the European Union's Horizon 2020 Research and Innovation program under Grant Agreement no. 654168. 
One of the authors received support through the ENIGMASS Labex (France).



\bibliographystyle{plain}
\bibliography{biblio}

\begin{thebibliography}{10}

\bibitem{AtlasPixels}
G.~Aad, M.~Ackers, F.A. Alberti, M.~Aleppo, G.~Alimonti, et~al.
\newblock {ATLAS} pixel detector electronics and sensors.
\newblock {\em JINST}, 3:P07007, 2008.

\bibitem{IBLTDR}
{ATLAS IBL Community}.
\newblock {ATLAS} insertable b-layer technical design report.
\newblock Technical report, CERN, 2010.

\bibitem{CMSPixels}
Aaron Dominguez.
\newblock {The CMS pixel detector}.
\newblock {\em Nucl. Instrum. Meth.}, A581:343--346, 2007.

\bibitem{HL-LHC}
S.~McMahon, P.~Allport, H.~Hayward, and B.~Di~Girolamo.
\newblock {Initial Design Report of the ITk: Initial Design Report of the ITk}.
\newblock Technical Report ATL-COM-UPGRADE-2014-029, CERN, Geneva, Oct 2014.

\bibitem{itk_strips_tdr}
ATLAS Collaboration.
\newblock {Technical Design Report for the ATLAS ITk Strip Detector}.
\newblock Technical Report ATL-COM-UPGRADE-2017-006, CERN, Geneva, Mar 2017.

\bibitem{bib:nim2012}
{M.~Bomben et al.}
\newblock {Development of Edgeless n-on-p Planar Pixel Sensors for future ATLAS
  Upgrades}.
\newblock {\em Nucl. Instr. and Meth. A}, 712:41--47, 2013.

\bibitem{FEI4}
{M.~Garcia-Sciveres et al.}
\newblock The {FE-I4} pixel readout integrated circuit.
\newblock {\em Nucl. Instr. and Meth. A}, 636:S155--S159, 2011.

\bibitem{bib:metal}
E.~Vianello et~al.
\newblock Optimization of double-side 3d detector technology for first
  productions at {FBK}.
\newblock In {\em Nuclear Science Symposium and Medical Imaging Conference
  (NSS/MIC), 2011 IEEE}, pages 523--528, 2011.

\bibitem{rossi2006pixel}
L.~Rossi, P.~Fischer, T.~Rohe, and N.~Wermes.
\newblock {\em Pixel detectors: From fundamentals to applications}.
\newblock Springer Science \& Business Media, 2006.

\bibitem{USBpix}
M.~Backhaus et. al.
\newblock Development of a versatile and modular test system for {ATLAS} hybrid
  pixel detectors.
\newblock {\em Nucl. Instr. Meth. A}, 650(1):37 -- 40, 2011.
\newblock International Workshop on Semiconductor Pixel Detectors for Particles
  and Imaging 2010.

\bibitem{Jansen2016}
H.~Jansen, S.~Spannagel, et~al.
\newblock Performance of the {EUDET}-type beam telescopes.
\newblock {\em EPJ Techniques and Instrumentation}, 3(1):7, 2016.

\bibitem{mimosa26}
C.~Hu-Guo et~al.
\newblock First reticule size maps with digital output and integrated zero
  suppression for the {EUDET-JRA1} beam telescope.
\newblock {\em Nucl. Instr. Meth. A}, 623(1):480 -- 482, 2010.
\newblock 1st International Conference on Technology and Instrumentation in
  Particle Physics.

\bibitem{RCE}
{\em {RCE}, the Reconfigurable Cluster Element,
  https://rceproject.web.cern.ch/}.

\bibitem{eutelescope}
{\em http://eutelescope.web.cern.ch/}.

\bibitem{Millipede}
V.~Blobel.
\newblock Millepede ii: Linear least squares fits with a large number of
  parameters.
\newblock In {\em Institut fur Experimentalphysik Universitat Hamburg 2007}.

\bibitem{root}
{\em ROOT Data Analysis Framework, https://root.cern.ch/}.

\bibitem{tbmon2}
{\em https://bitbucket.org/TBmon2/tbmon2/overview}.

\bibitem{stefano}
S.~Terzo.
\newblock {\em Development of radiation hard pixel modules employing planar
  n-in-p silicon sensors with active edges for the ATLAS detector at HL-LHC}.
\newblock PhD thesis, Technische Universitat Munchen, Max-Planck-Institut fur
  Physik, 2015.

\end{thebibliography}

\end{document}